\documentstyle[prd,aps,multicol,epsf]{revtex}
\widetext
\renewcommand{\narrowtext}{\begin{multicols}{2} \global\columnwidth20.5pc}
\renewcommand{\widetext}{\end{multicols} \global\columnwidth42.5pc}
\def\Lrule{\vspace*{-0.2in}\noindent\vrule width3.5in height.2pt
  depth.2pt \vrule depth0em height1em}
\def\Rrule{\vspace{-0.1in}\hfill\vrule depth1em height0pt \vrule
  width3.5in height.2pt depth.2pt\vspace*{-0.1in}}
\def\bml{\begin{mathletters}}
\def\eml{\end{mathletters}}
\def\beq{\begin{equation}}
\def\eeq{\end{equation}}
\def\bea{\begin{eqnarray}}
\def\eea{\end{eqnarray}}
\def\ba{\begin{array}}
\def\ea{\end{array}}

\def\to{\rightarrow}
\def\a{\alpha}

\def\g{\gamma}
\def\l{\lambda}
\def\m{\mu}
\def\n{\nu}
\def\z{\zeta}

\def\e{{\rm e}}
\def\tr{\,{\rm tr}\,}
\def\diag{\,{\rm diag}\,}
\def\re{{\rm Re}\,}
\def\pf{\,{\rm Pf}\,}
\begin{document}
\preprint{TIT-HEP-442}
\draft
\title{Massive chiral random matrix ensembles 
at ${\bf \beta}$ = 1 \& 4 :\\
QCD Dirac operator spectra}
\author{Taro Nagao}
\address{
Department of Physics, Graduate School of Science,
Osaka University,
Toyonaka, Osaka 560-0043, Japan
}
\author{Shinsuke M. Nishigaki}
\address{
Department of Physics, Faculty of Science,
Tokyo Institute of Technology,
Oh-okayama, Meguro, Tokyo 152-8551, Japan
}
\date{March 1, 2000}
\maketitle
\begin{abstract} 
The zero momentum sectors in effective theories of QCD
coupled to pseudoreal (two colors) and real (adjoint) quarks
have alternative descriptions in terms of 
chiral orthogonal and symplectic ensembles of random matrices.
Using this correspondence,
we compute correlation functions of Dirac operator eigenvalues
within a sector with an arbitrary topological charge
in a presence of finite quark masses of the order of 
the smallest Dirac eigenvalue.
These novel correlation functions,
expressed in terms of Pfaffians,
interpolate between
known results for the chiral and quenched limits
as quark masses vary.
\end{abstract}
\pacs{PACS number(s): 05.40.-a, 12.38.Aw, 12.38.Lg}
\renewcommand{\thefootnote}{\fnsymbol{footnote}}
\setcounter{footnote}{0}
\narrowtext
\section{introduction}
Random matrix theory of disordered Hamiltonians \cite{Meh,GMW}
relies upon an ansatz that in the ergodic regime where 
the system size $L$ is much larger than 
the elastic mean free path but much smaller than the 
localization length,
details of the Hamiltonian are lost except for its  
time-reversal and internal symmetries.
This ansatz was materialized by Efetov \cite{Efe}
who derived Wigner-Dyson statistics out of 
Anderson tight-binding model by retaining only
the zero-momentum mode of his spectral non-linear $\sigma$ model (NL$\sigma$M).

By taking the flavor symmetry among quarks 
into consideration 
as an additional internal symmetry,
Verbaarschot and collaborators \cite{ShV,VZ,Vrev} 
have reinterpreted this wisdom
in the context of quantum chromodynamics (QCD).
In this context,
the spectral NL$\sigma$M is a supersymmetric extension 
\cite{DOTV,TV}
of the conventional NL$\sigma$M over the coset
manifold associated with 
the 
chiral symmetry breaking \cite{LS,SmV}:
\bml
\begin{eqnarray}
&&Z^{(2)}(\theta; M) =
\int_{SU(N_f)} \!\!\!\!\!\!\!\!\!\!\!\!\!
dU\ \exp
(\re \tr \e^{i\theta/N_f} M U^\dagger ),\\
&&Z^{(1)}(\theta; M) =
\int_{SU(2N_f)/Sp(2N_f)} 
\!\!\!\!\!\!\!\!\!\!\!\!\!\!\!\!\!\!\!\!\!\!\!\!\!\!\!\!\!\!\!\!\!\!\!\!\!
dU\ 
\exp(
\re \tr\e^{i\theta/N_f}  M U {\bf J} U^T /2),\\
&&Z^{(4)}(\theta; M) =
\int_{SU(N_f)/SO(N_f)}  
\!\!\!\!\!\!\!\!\!\!\!\!\!\!\!\!\!\!\!\!\!\!\!\!\!\!\!\!\!\!\!\!\!\!
dU\  
\exp( \re  \tr\e^{i\theta/(N_{c}N_f)} M U U^T ),
\end{eqnarray}
\label{fvpftheta}
\eml
after retaining only the zero mode.
Here the superscripts (2, 1, 4) of Dyson indices $\beta$
refer to the anti-unitary symmetry of 
Euclidean Dirac operators (which are considered as
stochastic Hamiltonians)
in three classes of QCD \cite{V3foldway}:
\bea
\beta=2 &:&\ \,
N_c\geq 3,\ N_f\ \mbox{fundamental Dirac fermions}, \nonumber\\
\beta=1 &:&\ \, 
N_c= 2,\ N_f\ \mbox{fundamental Dirac fermions},\\
\beta=4 &:&\ \, 
N_c\geq 2,\ N_f\ \mbox{adjoint Majorana fermions}. \nonumber
\eea
The rescaled 
quark
mass matrices $M$ are defined as
\begin{eqnarray}
&&M = \diag(\m_1, \ldots, \m_{N_f})\ \ \ \ \ \ \ \ \ \ \ \ \,
 (\beta=2,4),
\nonumber\\
&&M = \diag(\m_1, \ldots, \m_{N_f})\otimes J\ \ \ \ \ \ \ 
 (\beta=1),
\\
&&
\mu_{i}\equiv  \Sigma L^4 m_{i},\ \ 
{\bf J}=\openone_{N_f} \otimes J,\ \ 
J=\left( 
\ba{cc}
 0 & 1 \\
-1 & 0
\ea
\right), 
\nonumber
\end{eqnarray}
with a limit
\beq
L\to\infty,\ \ m_i\to 0,\ \ \mu_{i}: \mbox{fixed}
\label{LSlim}
\eeq
being assumed,
$\Sigma$ stands for the quark condensate in the chiral limit,
and $\theta$ stands for the vacuum angle.
By the same token as the spectral NL$\sigma$M of the tight-binding model
was derived from a conventional random matrix ensembles \cite{VWZ},
these NL$\sigma$Ms have an alternative derivation from 
{\it chiral} random matrix ensembles ($\chi$RMEs) \cite{ShV,HV95}:
\bea
&&{Z}^{(\beta)}_\nu (\{m\})=\int_0^{2\pi}\frac{d\theta}{2\pi}
\e^{-i\n \theta}
Z^{(\beta)}(\theta; \{m\})\nonumber\\
&&\ \ \ \ =\int dW \e^{-\beta \tr V(W^\dagger W)}
\prod_{i=1}^{N_f}
\det 
\left(
\ba{cc}
m_i & W \\
-W^\dagger & m_i
\ea
\right),
\label{ZchiRME}
\eea
under a limit
\beq
N\to\infty,\ \ m_i\to 0,\ \ \mu_{i}= \pi \bar{\rho}(0) m_{i}: \mbox{fixed}.
\label{microlim}
\eeq
Here the integrals are over complex, real, and quaternion real 
$(N+\nu)\times N$ matrices $W$ for $\beta=2,1,4$, respectively,
and
$\bar{\rho}(0)$ stands for the large-$N$ 
spectral density of the random matrix
${\cal D}=\left({\ \ 0 \ \ \ \ W  \atop -W^\dagger\ \  0 }\right)$:
\beq
\bar{\rho}(\lambda)=\lim_{N\to\infty}
\langle \tr \delta(\lambda-i{\cal D}) \rangle,
\eeq
at the origin.
It is understood for $\beta=4$ that twofold degenerated eigenvalues
in the determinant are only counted once, and the topological charge
$\n$ is substituted by $\n N_c$.
These $\chi$RMEs are motivated by
the microscopic theories (Euclidean QCD) on a lattice, 
with a crude simplification of replacing
matrix elements of the anti-Hermitian Dirac operator
$/\!\!\!\!D= (\partial_\mu + i A_\mu)\gamma_\mu$ 
by random numbers ${\cal D}$ generated according to 
the weight $\e^{-\beta \tr V(W^\dagger W)}$.
Under this correspondence, the microscopic limit (\ref{microlim})
is equivalent to Leutwyler-Smilga limit (\ref{LSlim}),
since the size $N$ of the matrix $W$ is interpreted as
the number of cites $L^4$ of the lattice on which QCD is discretized,
and the Dirac spectral density at zero virtuality $\bar{\rho}(0)$
is related to the quark condensate by Banks-Casher relation
$\Sigma=\pi\bar{\rho}(0)/L^4$ \cite{BC}.
Using the $\chi$RME representation,
various correlation functions
of microscopically rescaled Dirac eigenvalues,
\beq
N\to\infty,\ \ \lambda \to 0,\ \ 
\zeta= \pi \bar{\rho}(0) \lambda: \mbox{fixed},
\label{microlim_lambda}
\eeq
have been computed 
for all three values of $\beta$ in
the massless case $\mu\equiv 0$
\cite{NSl,For,AST,SlN,FH,TW,V_chGOE,NF95,NF98}.
On the other hand, in a presence of finite $\mu$'s,
Dirac eigenvalues 
comparable to $\mu$'s
are expected to obey statistics that interpolate
the chiral ($\mu\to 0$) and quenched limits ($\mu\to \infty$ or $N_f=0$).
These novel spectral correlation functions have 
been analytically computed, until recently, solely for the chiral unitary 
($\beta=2$) ensemble \cite{GW,JSV,DN98a,WGW,JNZ,NDW}.
Therefore we aim to treat the remaining cases, 
chiral orthogonal ($\beta=1$) and 
symplectic ($\beta=4$) ensembles with finite mass parameters, 
and compute 
Dirac 
eigenvalue correlation functions for these ensembles.
We anticipate that advances in numerical simulations of 
lattice QCD with dynamical quarks
\cite{BMW,FFHLS} will confirm our
analytical results, Eqs.(\ref{Pf1}) and (\ref{Pf4}) below,
in a foreseeable future.

This Article is organized as follows. 
In Sect.\ 2 we compute the correlation functions for 
the chiral orthogonal ensemble, by utilizing the 
quaternion determinant method
developed in Ref.\cite{NF99}.
In Sect.\ 3 we exhibit 
an
explicit form of
the correlation functions for the chiral symplectic ensemble,
which was obtained by the Authors \cite{NN} 
as a corollary to the computation of the partition function.
In Appendix A we collect definitions related to a quaternion determinant.
In Appendices B and C we present alternative expressions 
of the quaternion kernels for the orthogonal ensemble, 
and the symplectic ensembles with quadruply degenerated masses,
respectively. It enables us to identify our results with
those in a paper by Akemann and Kanzieper \cite{AK},
which appeared after the 
Letter by the present Authors \cite{NN}
and computed
a 1-level correlator with a single mass (for $\beta=1$) and
$p$-level correlators with quadruply degenerate masses (for $\beta=4$).
\section{orthogonal ensemble}
We start by expressing the partition function (\ref{ZchiRME})
of the $\chi$RME in terms of eigenvalues $x_i=\lambda_i^2$
of the Wishart matrix $W^\dagger W$
(up to an overall constant): 
\begin{eqnarray}
&&{Z}^{(\beta)}_\nu (\{ m \})=
\biggl(\prod_{i=1}^\alpha m_i^\nu \biggl)
\Xi_0
(\{ m \}),
\nonumber\\
&&\Xi_0
(\{ m \})=\frac{1}{N!}
\int_0^\infty 
\!\!\!\!\cdots \int_0^\infty 
\prod_{j=1}^N dx_j\, P(\{x\};\{m\}) , \\
&&P(\{x\};\{m\})=
\prod_{j=1}^N
\biggl(
{\rm e}^{-\beta V(x_j)} x_j^{{\beta}\frac{\nu+1}{2} -1} 
\prod_{i=1}^\a (x_j+m_i^2)
\biggr)\nonumber\\ 
&&
\ \ \ \ \ \ \ \ \ \ \ \ \ \ \ \ \ \ 
\times\prod_{j>k}^N |x_j-x_k|^\beta .
\end{eqnarray}
The indices $\beta$ and $\nu$ are suppressed for simplicity.
Since the partition function (\ref{ZchiRME}) is even under $\nu\mapsto-\nu$,
we have set $\nu$ non-negative integer, without loss of generality.
The $p$-level correlation function of
the Wishart matrix $W^\dagger W$ is defined as
\bea
&&\sigma(x_1,\ldots,x_p;\{m\})=
\frac{\Xi_p (x_1,\ldots,x_p;\{ m \})}{\Xi_0 (\{ m \})},
\label{sigma_p}
\\
&&\Xi_p (x_1,\ldots,x_p;\{ m \})=
\nonumber\\
&&\frac{1}{(N-p)!}
\int_0^\infty \!\!\!\!\cdots \int_0^\infty 
\prod_{j=p+1}^N dx_j \,
P(\{x\};\{m\}).
\label{Xi_p}
\eea
Then the $p$-level correlation function of 
the block off-diagonal Hermitian matrix $i{\cal D}$,
\beq
\rho(\l_1,\ldots,\l_p;\{m\})=
\langle \prod_{k=1}^p {\rm tr}\,\delta(\lambda_k-i{\cal D}) \rangle,
\eeq
is expressed in terms of
$\sigma$ multiplied by the Jacobian 
of the transformation $\lambda\mapsto x=\lambda^2$ :
\beq
\rho(\l_1,\ldots,\l_p;\{m\})=
2^p 
\prod_{j=1}^p |\l_j|\,
\sigma(\l_1^2,\ldots,\l_p^2;\{m\}).
\label{rhosigma}
\eeq
As the universality of correlation functions of
the unitary ensemble in the microscopic limit
(\ref{microlim_lambda}) \cite{Nis,ADMN,KF,DN98a}
is known to inherit to
those of orthogonal and symplectic ensembles \cite{Wid,SV},
it suffices to concentrate on Laguerre ensembles, $V(x)= x$.
This leads to Wigner's semi-circle law
\beq
\bar{\rho}(\lambda)=\frac{2}{\pi}\sqrt{2N-\lambda^2}.
\eeq

Now we concentrate on $\beta=1$, 
and define new variables $z_j$ as 
\beq
z_j =
\left\{
\ba{rl}
-m^2_j \ \ (\leq 0), & \ \ j=1,\ldots,\alpha,  \\ 
x_{j-\alpha}\ \ (\geq 0), & \ \ j=\alpha+1,\ldots,\alpha+N.
\ea
\right.
\label{zj}
\eeq
Then the multiple integral (\ref{Xi_p}) is
expressed as
\bea
&&
\Xi_p(z_1,\ldots,z_{\a+p}) \!= \!
\frac{1}{\prod_{j=1}^{\alpha}\! \sqrt{w(z_j)} 
\prod_{j > k}^{\alpha} \!\mid z_j - z_k \mid} \! \times 
\label{Xi_p_z}\\
&&
\frac{1}{(N-p)!}
\int_0^\infty\!\!\!\!\cdots\int_0^\infty 
\!\!\!\!\prod_{j=\a+p+1}^{\a+N} \!dz_j \!
\prod_{j=1}^{\alpha + N}\! \sqrt{w(z_j)} \!
\prod_{j > k}^{\alpha + N}\! \mid z_j - z_k \mid, \! 
\nonumber
\eea
where
\begin{equation}
w(z) = |z|^{\nu-1} {\rm e}^{-2z}.
\label{Laguerreweight}
\end{equation}
Eq.(\ref{Xi_p_z}) resembles an 
$(\alpha + p)$-level correlation function of 
the conventional (massless) Laguerre ensemble 
with $\alpha + N$ levels. However, conventionally the levels 
$z_1,\ldots,z_{\alpha+p}$ are all positive, 
while in the present case some of 
them ($z_1,\ldots,z_{\alpha}$) are negative. 
We carefully incorporate 
this fact into the following evaluation.

Let us denote 
the integrand in Eq.(\ref{Xi_p_z}) as
\begin{equation}
p(z_1,\ldots,z_\g) = 
\prod_{j=1}^{\g} {\sqrt{w(z_j)}} 
\prod_{j > k}^{\g} \mid z_j - z_k \mid,
\end{equation}
with $\gamma\equiv \alpha+N$.
It can be readily seen that 
\bea
&&p(z_1,\ldots,z_\g) = 
\prod_{j=1}^{\g} 
{\sqrt{w(z_j)}} \prod_{j>k}^{\g} (z_j - z_k) \times \\
&&
\left\{
\ba{ll}
{\rm Pf}[{\rm sgn}(z_k-z_j)]_{j,k = 1,\ldots,\g}
&\ \ (\g: \mbox{even}) \\
{\rm Pf}
  \left[ 
  \begin{array}{ll}
    [{\rm sgn}(z_k-z_j)]_{ j,k = 1,\ldots,\g}
& [g_j]_{j=1,\ldots,\g} \\ 
\left[ - g_k \right]_{k=1,\ldots,\g} & 0 
  \end{array} 
  \right]
&\ \  (\g: \mbox{odd}) ,
\ea
\right.
\nonumber
\end{eqnarray}
with $g_j=g_k=1$.
The Pfaffians in the above can be represented as
quaternion determinants \cite{DysonQ,MehtaQ,MM,NF95,NF98,NF99,FNH}.
In doing so, we need to introduce monic skew-orthogonal polynomials
$R_n(z)= z^n + \cdots$,
which satisfy the skew-orthogonality relation:
\beq
\langle R_{2n}, R_{2m+1} \rangle_R =
 - \langle R_{2m+1}, R_{2n}\rangle_R = r_n \delta_{nm}, \; 
\mbox{others}=0,\! 
\end{equation}
where
\beq
\langle f, g \rangle_R
=
\!\!\int_{0}^{\infty} \!\!\!dz 
\sqrt{w(z)} 
g(z) 
\!\int_{0}^{z} \!{dz'} 
\sqrt{w(z')}
f(z') 
 - (f\leftrightarrow g),
\eeq
and (integrated) `wave functions',
\bml
\bea
&&\Psi_n(z) = \sqrt{w(z)} R_n(z), \\
&&\Phi_n(z) = \int_0^{\infty}{dz'} \,{\rm sgn}(z-z') \sqrt{w(z')} R_n(z').
\eea
\label{PsiPhi}
\eml
Note that for negative $z$, $\Phi_n(z)$
is a constant:
\beq
\Phi_n(z< 0)=\Phi_n(0)\equiv -s_n .
\label{sn}
\eeq

Now we present the following theorems:
\par
\bigskip
\noindent
{\em Theorem 1}
\par
\bigskip
\noindent
For even $\g$, we can express $p(z_1,\ldots,z_{\g})$ as 
\beq
p(z_1,\cdots,z_{\g}) = 
\biggl( \prod_{j=0}^{\g/2-1} r_{j} 
\biggr) {\rm Tdet} [f(z_j,z_k)]_{j,k = 1,\ldots,\g}.
\eeq
The quaternion elements $f(z,z')$ are represented as
\begin{equation}
f(z,z')= \left[ \begin{array}{cc} S(z,z') & I(z,z') \\ 
D(z,z') & S(z',z) \end{array} \right].
\end{equation}
The functions $S(z,z')$, $D(z,z')$ and $I(z,z')$ are given by
\bml
\bea
S(z,z')\! &=&\!\!\! \sum_{n=0}^{\g/2-1}\!\!
\frac{
\Phi_{2n}(z) \Psi_{2n+1}(z')\! -\! \Phi_{2n+1}(z) 
\Psi_{2n}(z')}{r_{n}},\!\!\!\!\\
D(z,z')\! &=&\!\!\! \sum_{n=0}^{\g/2-1}\!\!
\frac{
\Psi_{2n}(z) \Psi_{2n+1}(z') \!-\! \Psi_{2n+1}(z) 
\Psi_{2n}(z')}{r_{n}},\!\!\!\!\\
I(z,z')\!&=&\!\!\! \sum_{n=0}^{\g/2-1}\!\!
\frac{
\Phi_{2n}(z) \Phi_{2n+1}(z') \!-\! \Phi_{2n+1}(z) 
\Phi_{2n}(z')}{r_{n}}.\!\!\!\!
\eea
\eml
\par
\bigskip
\noindent
{\em Theorem 2}
\par
\bigskip
\noindent
For odd $\g$, we can express $p(z_1,\ldots,z_{\g})$ as 
\bea
&&
p(z_1,\ldots,z_{\g}) = \\
&&
\biggl( \prod_{j=0}^{[\g/2]-1} r_{j} \biggr) 
s_{\g-1} {\rm Tdet} 
[{f^{\rm odd}}(z_j,z_k)]_{j,k = 1,\ldots,\g}. 
\nonumber
\eea
The quaternion elements are represented as
\begin{equation}
{f^{\rm odd}}(z,z') = \left[ \begin{array}{cc} 
{S^{\rm odd}}(z,z') 
& {I^{\rm odd}}(z,z') \\ 
{D^{\rm odd}}(z,z') & {S^{\rm odd}}(z',z) \end{array} \right]
\end{equation}
and $s_n$ is defined in Eq.(\ref{sn}).
The functions ${S^{\rm odd}}$, ${D^{\rm odd}}$, and ${I^{\rm odd}}$ are 
given in terms of $S$, $D$, and $I$ in {\em Theorem 1} according to
\bml
\begin{eqnarray}
{S^{\rm odd}}(z,z') & = & S(z,z') \Big|_* 
+ {{\Psi}_{\g-1}(z') \over s_{\g-1}},\\
{D^{\rm odd}}(z,z') & = & D(z,z') \Big|_*,\\
{I^{\rm odd}}(z,z')  & = & I(z,z') \Big|_* + 
{{\Phi}_{\g-1}(z) - {\Phi}_{\g-1}(z') \over s_{\g-1}} .  
\end{eqnarray}
\eml
Here $*$ stands for a substitution
\beq
{R}_n(z) \mapsto 
R_n(z)-\frac{s_n}{s_{\g-1}}R_{\g-1}(z)
\eeq
for $n=0,\ldots,\g-2$,
associated with a change in the upper limit of the sum
\beq
\g/2 -1 \mapsto [\g/2]-1.
\eeq
\par
\bigskip
\noindent
{\em Theorem 3}
\par
\bigskip
\noindent
Let the quaternion elements $q_{jk}$ of a selfdual $n \times n$ matrix 
$Q_n$ 
depend on $n$ real or complex variables $z_1,\cdots,z_n$ as 
\begin{equation} 
q_{jk} = f(z_j,z_k). 
\end{equation}
We assume that $f(z,z')$ satisfies the following conditions.
\bea
&&\int \!f(z,z) d\mu(z) = c, \\
&&
\int \!f(z,z'') f(z'',z') d\mu(z'') \!=\! f(z,z') \!+ \! 
 \lambda f(z,z')\!- \!f(z,z') \lambda.\!
 \nonumber
\eea
Here $d\mu(z)$ is a suitable measure, $c$ is a 
constant scalar, and $\lambda$ is a constant quaternion. Then we 
have
\begin{equation} 
\int {\rm Tdet}\, Q_n \, d\mu(z_n) = 
(c-n+1) {\rm Tdet}\, Q_{n-1},
\end{equation}
where $Q_{n-1}$ 
is the $(n-1) \times (n-1)$ matrix obtained by removing the row 
and the column which contain $z_n$.
It is straightforward to show that the quaternion element $f(z,z')$ and 
${f^{\rm odd}}(z,z')$ in {\em Theorem 1} and {\em Theorem 2} 
both satisfy the conditions imposed on $f(z,z')$ in {\em Theorem 3}
with $d\mu(z)=w(z) dz$. 
This means that we can write
\bea
&&\Xi_p(z_1,\ldots,z_{\alpha+p}) \!=\! 
\frac{\prod_{j=0}^{[(\alpha+N)/2]-1} r_{j} }{\prod_{j=1}^{\alpha} 
\!\sqrt{w(z_j)} 
\prod_{j > k}^{\alpha} \!\mid z_j - z_k \mid}
\!\times\! \\
&&
\left\{
\ba{lll}
           &{\rm Tdet}[f(z_j,z_k)]_{j,k=1,\ldots,\alpha+p}
& (\alpha+N : \mbox{even})\\
s_{\a+N-1}\!\!\! & {\rm Tdet}[{f^{\rm odd}}(z_j,z_k)]_{j,k=1,\ldots,\alpha+p}
& (\alpha+N: \mbox{odd})
\ea
\right. \! .\!
\nonumber
\eea
Since the final result in the asymptotic limit $N \rightarrow \infty$ 
should be insensitive to the parity of $N$, we consider only even 
$\alpha +N$ henceforth. Then the $p$-level correlation function 
(\ref{sigma_p})
is finally written as
\beq
\sigma(x_1,\ldots,x_p;m_1,\ldots,m_{\alpha})= 
\frac{{\rm Tdet}[f(z_j,z_k)]_{j,k=1,\ldots,\alpha+p}}%
{{\rm Tdet}[f(z_j,z_k)]_{j,k=1,\ldots,\alpha}}.
\label{sigma}
\eeq

Now we proceed to take the asymptotic limit of the correlation function,
by making use of explicit forms for the skew-orthogonal polynomials
and their norms associated with the weight (\ref{Laguerreweight}) 
obtained by Nagao and Wadati \cite{NW}:
\begin{eqnarray}
R_{2n}(z) & = & - \frac{(2n)!}{2^{2n+1}} 
\frac{d}{dz} L_{2n+1}^{(\nu-1)}(2z) ,
 \nonumber \\
R_{2n+1}(z) & = & - \frac{(2n+1)!}{2^{2n+1}} L_{2n+1}^{(\nu-1)}(2z) \\
&&-\frac{(2n)!}{2^{2n+2}}(2n+\nu)\frac{d}{dz}L_{2n}^{(\nu-1)}(2z) , 
\nonumber\\
r_n &=& 2^{ - 4 n - \nu} (2 n)! (2 n +  \nu)! \ ,\nonumber
\end{eqnarray}
expressed in terms of the Laguerre polynomials
\begin{equation}
L_n^{(a)}(z) = \frac{z^{-a} {\rm e}^{z}}{n!}
\frac{d^n}{dz^n}({\rm e}^{- z} z^{n + a}).
\end{equation}
We need to evaluate the local asymptotics of 
the quaternion function $f(z,z')$, 
whose constituent Laguerre polynomials tend to 
\begin{equation}
L_n^{(a)}(z) \sim \left\{ 
\begin{array}{ll} 
\left( {n}/{z} \right)^{{a}/{2}} 
J_a(2 \sqrt{n z}) & \ \ (z>0) \\   
\left( {n}/{|z|} \right)^{{a}/{2}}   
I_a(2 \sqrt{n |z|}) & \ \ (z<0)
\end{array} 
\right. , 
\label{LJI}
\end{equation}
as $n\to \infty$, $z\to 0$, with $n z:$ fixed.
Three cases should be considered separately: 
\[
(a)\ z, z'>0,\ \ \ \ (b)\ z<0,\ z'>0,\ \ \ \ (c)\ z, z'<0
\] 
(the case $z>0$, $z'<0$ is unnecessary because of the selfduality 
$f(z,z') = {\hat f}(z',z)$). 
\par
\bigskip
\noindent
$(a)$ $z, z'>0$
\par
\bigskip
\noindent
We define microscopic variables $\zeta$ and $\zeta'$ as
\begin{equation}
z = \frac{\zeta^2}{8 N}, \ \ \ z' = \frac{{\zeta'}^2}{8 N},
\label{zz}
\end{equation}
according to Eq.(\ref{microlim_lambda}) with
$\pi \bar{\rho}(0) = 2\sqrt{2N}$.
If both arguments are positive, the asymptotic limit is 
known to be \cite{NF95} 
\widetext
\Lrule
\bml
\bea
&&S_{++}(\zeta, \zeta')\equiv\frac{1}{8N}S(z,z')
\sim  \frac14 \int_0^1 dt\, {t}^2  
 \int_0^{\z} ds \left( s\, J_{\nu-1}({t} s) 
\frac{J_\nu ({t}\z')}{\z'} -  J_{\nu}({t} s) 
J_{\nu-1}({t}\z') \right) +  \frac{J_\nu (\z')}{4\z'} , 
\label{Szz}\\
&&D_{++}(\zeta, \zeta')\equiv
\frac{1}{(8N)^2}
D(z,z')
\sim  \frac{1}{16} \int_0^1 dt\, {t}^2 
\bigl(
J_{\nu-1}({t}\z) \frac{J_{\nu}({t}\z')}{\z'} 
- \frac{J_{\nu}({t}\z)}{\z} J_{\nu-1}({t}\z')
\bigr) ,
\label{Dzz}\\  
&&I_{++}(\zeta, \zeta')\equiv
I(z,z')
\sim - 
\int_0^1 dt\, {t}^2 
\int_0^{\z} du \int_0^{\z'} dv
\bigl( 
u\, J_{\nu-1}({t} u)J_{\nu}({t} v) 
- J_{\nu}({t} u)v\, J_{\nu-1}({t} v)
\bigr)  
- \int_{\z}^{\z'} J_{\nu}(u) du
- {\rm sgn}(\z - \z').
 \nonumber\\
&& \label{Izz}
\end{eqnarray}
\eml
\par
\bigskip
\noindent
$(b)$ $z<0, z'>0$
\par
\bigskip
\noindent
We define microscopic variables $\mu$ and $\zeta$ as
\begin{equation}
z = -\frac{\m^2}{8 N}, \ \ \ z' = \frac{\zeta^2}{8 N},
\label{mz}
\end{equation}
according to Eqs.(\ref{microlim}) and (\ref{microlim_lambda}).
When one of the arguments is negative, 
the identity (\ref{sn})
should be taken into consideration.  We find
\bml
\bea
&&S_{-+}(\m, \zeta)
=S_{+}(\zeta) 
\equiv
\frac{1}{8N}S(z,z')
\sim \frac{J_{\nu}(\z)}{4\z} ,
\label{43a}
\\
&&S_{+-}(\z, \m)\equiv
\frac{1}{8N}S(z',z)
\sim  \frac14
\int_0^1 dt\, {t}^2  \int_0^{\z}ds
\left(
s\,J_{\nu-1}({t}s) \frac{I_\n ({t}\m)}{\m} 
- J_{\nu}({t}s) I_{\nu-1}({t}\m)
\right)
+ \frac{I_{\nu}(\m)}{4\m}  , \\
&&D_{-+}(\m, \z)\equiv
\frac{1}{(8N)^2}D(z,z')
\sim  \frac{1}{16} \int_0^1 dt\, {t}^2 
\left(
I_{\nu-1}({t}\m) \frac{J_{\nu}({t}\z)}{\z}  - 
\frac{I_{\nu}({t}\m)}{\m} J_{\nu-1}({t}\z)
\right),  \\
&&I_{-+}(\m, \z)=I_{+}(\zeta)\equiv
I(z,z')\sim - \int_0^{\z}  J_{\nu}(u) du+ 1 .
\label{43d}
\eea
\eml
\par
\bigskip
\noindent
$(c)$ $z, z'<0$
\par
\bigskip
\noindent
We define microscopic variables $\mu$ and $\zeta$ as
\begin{equation}
z = -\frac{\m^2}{8 N}, \ \ \ z' = -\frac{{\m'}^2}{8 N}.
\label{mm}
\end{equation}
We can readily derive
\bml
\bea
&&S_{--}(\m, \m')
=S_{-}(\m')
\equiv
\frac{1}{8N}S(z,z') 
\sim  \frac{I_{\nu}(\m')}{4\m'} ,
\label{45a}\\
&&D_{--}(\m, \m')\equiv
\frac{1}{(8N)^2}D(z,z')\sim 
 \frac{1}{16} \int_0^1 dt \,{t}^2
\left( I_{\nu-1}({t}\m) 
\frac{I_{\nu}({t}\m')}{\m'} - 
\frac{I_{\nu}({t}\m)}{\m} I_{\nu-1}({t}\m')
\right) ,  
\\
&&I_{--}(\m, \m')\equiv I(z,z')
 \sim  {\rm sgn}(\m - \m').
\eea
\eml
In Eqs.(\ref{43a}), (\ref{43d}), and (\ref{45a}),
we have introduced symbols with one sign subscript (e.g. $S_{+}$)
in order to indicate that they depend only on the second
arguments of those with two sign subscripts (e.g. $S_{-+}$).

The scaled correlation function $\rho_s$ is defined as
\beq
\rho_s(\zeta_1,\ldots,\zeta_p; \m_1,\ldots,\m_\alpha)=
\frac{1}{(8N)^p}
\rho(
\frac{\zeta_1}{\sqrt{8N}},\ldots,
\frac{\zeta_p}{\sqrt{8N}}; 
\frac{\mu_1}{\sqrt{8N}},\ldots,
\frac{\m_\a}{\sqrt{8N}}) .
\label{rhos}
\eeq
By making use of Eqs.(\ref{rhosigma}), (\ref{sigma}), and Dyson's equality
(\ref{Dysoneq}), we finally obtain
\bml
\bea
\rho_s(\zeta_1,\ldots,\zeta_p; \m_1,\ldots,\m_\alpha)
&&=
(-1)^{p(p-1)/2}
{2^p}
\prod_{k=1}^p |\zeta_k |
\frac{\pf
\left[
\ba{rrrr}
-I_{--}  & S_{--} & -I_{-+}  & S_{-+}  \\
-S^T_{--} & D_{--} & -S^T_{+-} & D_{-+}  \\
I^T_{-+}  & S_{+-} & -I_{++}  & S_{++}  \\
-S^T_{-+} & -D^T_{-+} & -S^T_{++} & D_{++}  
\ea
\right]}{
\pf\left[
\ba{rr}
 -I_{--}  & S_{--}  \\
-S^T_{--} & D_{--}  
\ea
\right]}
\nonumber\\
&&
=
(-1)^{p(p-1)/2}
{2^p}
\prod_{k=1}^p |\zeta_k | 
\frac{\pf
\left[
\ba{rrr}
D_{--} & -S^T_{+-} & D_{-+}  \\
S_{+-} & -I_{++}  & S_{++}  \\
-D^T_{-+} & -S^T_{++} & D_{++}  
\ea
\right]}{
\pf\left[
D_{--}
\right]}
\ \ \ \ \ \ \ \ \ \ \ \ \ \ \ 
(\a : \mbox{even})
\label{Pf1even}\\
&&
=
(-1)^{p(p-1)/2}
{2^p}
\prod_{k=1}^p |\zeta_k |
\frac{\pf
\left[
\ba{rrrr}
0~~  & S_- ~\; & -I_{+} ~\;& S_{+} ~\; \\
-S^T_{-} & D_{--} & -S^T_{+-} & D_{-+}  \\
I^T_{+}  & S_{+-} & -I_{++}  & S_{++}  \\
-S^T_{+} & -D^T_{-+} & -S^T_{++} & D_{++}  
\ea
\right]}{
\pf\left[
\ba{rr}
0~~  & S_{-} ~\; \\
-S^T_{-} & D_{--}  
\ea
\right]}
\ \ \ \ \ \ \; 
(\a : \mbox{odd}).
\label{Pf1odd}
\eea
\label{Pf1}
\eml
\noindent
The elements of the matrices 
$S_{\epsilon\epsilon'}$, $D_{\epsilon\epsilon'}$, 
$I_{\epsilon\epsilon'}$ 
and the row vectors 
$S_\epsilon$, $I_\epsilon$ 
($\epsilon, \epsilon'=+,-$)
in the above are defined as
\beq
(S_{++})_{k\ell}=S_{++}(\z_k, \z_\ell),\  
(S_{-+})_{i\ell}=S_{-+}(\m_i, \z_\ell),\ 
(S_{--})_{ij}=S_{--}(\m_i, \m_j),\ 
(S_{+})_{\ell}=S_{+}(\z_\ell),\ 
(S_{-})_{j}=S_{-}(\m_j),\ \mbox{etc.},
\eeq
where the subscripts take their values in
$i,j=1,\ldots,\a$ and $k,\ell=1,\ldots,p$.
In the last two lines we have exploited a Pfaffian identity
that holds for antisymmetric matrices $A$, $B$,
and a row vector $v$:
\bml
\bea
{\rm Pf}
\left[
\ba{c|c}
A &  
\ba{c}
v\\
\vdots\\
v
\ea
\\
\hline
-v^T \cdots -v^T & B
\ea
\right]
&=&{\rm Pf}[A] \,{\rm Pf}[B]
~~~~~~~~~~~~~~~~~~~~~~~~~~~~~~~~~~~~~~~~\; 
({\rm rank}(A), {\rm rank}(B) : \mbox{even}) \\
&=&
{\rm Pf}
\left[
\ba{c|c}
A &  
\ba{c}
1\\
\vdots\\
1
\ea
\\
\hline
-1 \cdots -1 & 0
\ea
\right]
{\rm Pf}
\left[
\ba{cc}
0 & v\\
-v^T & B
\ea
\right]
~~~~~~~~ \,
({\rm rank}(A), {\rm rank}(B) : \mbox{odd}) .
\eea
\eml
\noindent
In a special case $p = \alpha = 1$, the expression (\ref{Pf1odd})
reduces to Akemann and Kanzieper's recent result \cite{AK}
(see Appendix B).

In the quenched limit $\mu_1,\ldots,\mu_\a\to \infty$
when the ratio of two Pfaffians is replaced by a minor 
$\pf 
\left[{
-I_{++} \ \  S_{++}  \atop
-S_{++}^T\ \ D_{++} }\right]
$,
the correlation function tends
to that of Laguerre orthogonal ensemble
computed by Nagao and Forrester,
Eqs.(2.21), (2.18), (2.19), and (3.20) in Ref.\cite{NF95},
with $\nu=2a+1$.
By the same token, it satisfies a sequence
\beq
\rho_s(\{\z\};\mu_1,\ldots,\mu_{\alpha})
\stackrel{\mu_{\alpha} \to \infty}{\longrightarrow}
\rho_s(\{\z\};\mu_1,\ldots,\mu_{\alpha-1})
\stackrel{\mu_{\alpha-1} \to \infty}{\longrightarrow}
\rho_s(\{\z\};\mu_1,\ldots,\mu_{\alpha-2})
\stackrel{\mu_{\alpha-2} \to \infty}{\longrightarrow}
\cdots ,
\label{decouple}
\eeq
as each of the masses are decoupled
by sending to infinity.
To illustrate this decoupling, we exhibit in Fig.1 a plot of
the spectral density $\rho_s(\z;\m)$ ($\n=0, p=1, \a=1$)
that interpolates between known results for
the chiral
and quenched limits.
\section{symplectic ensemble}
Although correlation functions of the massive chiral
symplectic ensemble have previously been computed 
by the Authors \cite{NN},
we nevertheless present their explicit expressions
for the sake of completeness.
We concentrate on the case with an even $N_f(\equiv 2\a)$
number of flavors and pairwise degenerated mass parameters,
corresponding to adjoint Dirac fermions in the QCD context.
The scaled $p$-level correlation functions,
defined in Eq.(\ref{rhos}), is expressed
by construction as a ratio
of partition functions with $2\a$ and $2\a+4p$ flavors \cite{Dam,AD98},
\beq
\rho_s(\zeta_1,\ldots,\zeta_p;\{ \m \})=
C_{\a,\n}^{(p)}
\prod_{k>\ell}^p(\z_k^2-\z_\ell^2)^4
\prod_{k=1}^p
\Bigl(|\zeta_k|^3 \prod_{i=1}^\a (\zeta_k^2+\m_i^2)^2 
\Bigr)
\frac{Z^{(4)}_\nu(
\m_1,\m_1,\ldots,\m_\a,\m_\a,
\overbrace{i\z_1,\ldots,i\z_1}^{4},\ldots, 
\overbrace{i\z_p,\ldots,i\z_p}^{4})}{
Z^{(4)}_\nu(\m_1,\m_1,\ldots,\m_\a,\m_\a)}.
\eeq
Here $C_{\a,\n}^{(p)}$ stands for a constant to be fixed below.
Using an explicit form of the partition function,
Eq.(31) of Ref.\cite{NN}, and taking confluent limits
in $\z_k$'s, we obtain
\bml
\bea
\rho_s(\zeta_1,\ldots,\zeta_p; \{\m\})&=&
(-1)^{p(p+1)/2}
2^p
\prod_{k=1}^p |\zeta_k| 
\frac{\pf
\left[
\ba{rrr}
  I_{--}  & I_{-+} & S_{-+} \\
-I_{-+}^T & I_{++} & S_{++} \\
-S_{-+}^T & -S_{++}^T & D_{++}
\ea
\right]}{\pf[I_{--}]}
\ \ \ \ \ \ \ \ \ \ \ \ \ \ \ \ \ (\a : \mbox{even})
\label{Pf4even}\\
&=&
(-1)^{p(p+1)/2}
2^p
\prod_{k=1}^p |\zeta_k |
\frac{\pf
\left[
\ba{rrrr}
  I_{--}  & Q_- & I_{-+} & S_{-+} \\
-Q_-^T ~& 0~~ & -Q_+^T ~ & -P_+^T ~ \, \\
-I_{-+}^T & Q_+ & I_{++} & S_{++} \\
-S_{-+}^T & P_+ & -S_{++}^T & D_{++}
\ea
\right]}{\pf\left[
\ba{rr}
 I_{--}  & Q_-  \\
-Q_-^T ~ & 0~~ 
\ea
\right]}
\ \ \ \ \ \ \ \ \ \, (\a : \mbox{odd}).
\eea
\label{Pf4}
\eml
The elements of the matrices 
$S_{\epsilon\epsilon'}$, $D_{\epsilon\epsilon'}$, 
$I_{\epsilon\epsilon'}$
 and the column vectors 
$Q_\epsilon$, $P_\epsilon$ 
($\epsilon, \epsilon'=+,-$)
in the above are defined as
\begin{eqnarray}
&&(I_{--})_{ij}=
I_{--}(\m_i, \m_j)\equiv
\mu_i\mu_j
\int_0^1 dt\,t \int_0^1 du
\bigl(
I_{2\nu}(2{t}\mu_i) I_{2\nu}(2{t}u\mu_j)
-I_{2\nu}(2{t}u\mu_i) I_{2\nu}(2{t}\mu_j)
\bigr) ,\nonumber\\
&&(I_{-+})_{i\ell}=
I_{-+}(\m_i, \z_\ell)\equiv
\mu_i\z_\ell
\int_0^1 dt\,t \int_0^1 du
\bigl(
I_{2\nu}(2{t}\mu_i)J_{2\nu}(2{t}u\z_\ell) 
- I_{2\nu}(2{t}u\mu_i)J_{2\nu}(2{t}\z_\ell)
\bigr), \nonumber\\
&&(S_{-+})_{i\ell}=
S_{-+}(\m_i, \z_\ell)\equiv
\mu_i
\int_0^1 dt\,t^2 \int_0^1 du
\bigl(
I_{2\nu}(2{t}\mu_i) u\,J_{2\nu+1}(2{t}u\z_\ell)
-I_{2\nu}(2{t}u\mu_i) J_{2\nu+1}(2{t}\z_\ell) 
\bigr), \nonumber\\
&&(I_{++})_{k\ell}=
I_{++}(\z_k, \z_\ell)\equiv
\z_k\z_\ell
\int_0^1 dt\,t \int_0^1 du
\bigl(
J_{2\nu}(2{t}\z_k) J_{2\nu}(2{t}u\z_\ell)
-J_{2\nu}(2{t}u\z_k) J_{2\nu}(2{t}\z_\ell) 
\bigr),
\label{SDI}\\ 
&&(S_{++})_{k\ell}=
S_{++}(\z_k, \z_\ell)\equiv
{\z_k}
\int_0^1 dt\,t^2 \int_0^1 du
\bigl(
J_{2\nu}(2{t}\z_k)u\,J_{2\nu+1}(2{t}u\z_\ell)
-  J_{2\nu}(2{t}u\z_k) J_{2\nu+1}(2{t}\z_\ell) 
\bigr),\nonumber \\
&&(D_{++})_{k\ell}=
D_{++}(\z_k, \z_\ell)\equiv
\int_0^1 dt\,t^3 \int_0^1 du\,u
\bigl(
J_{2\nu+1}(2{t}\z_k)J_{2\nu+1}(2{t}u\z_\ell)
-J_{2\nu+1}(2{t}u\z_k)J_{2\nu+1}(2{t}\z_\ell)
\bigr),\nonumber\\
&&(Q_-)_j=\mu_j\int_0^1 dt \,{I_{2\nu}(2t\mu_j)},\ \ \ 
(Q_+)_\ell=\z_\ell\int_0^1 dt\, {J_{2\nu}(2t\z_\ell)},\ \ \ 
(P_+)_\ell=\int_0^1 dt \,t \,{J_{2\nu+1}(2t\z_\ell)},\nonumber
\end{eqnarray}
\Rrule
\narrowtext
\noindent
where the subscripts take their values in
$i,j=1,\ldots,\a$ and $k,\ell=1,\ldots,p$.
The overall constant is determined as the above by requiring
that in the quenched limit $\mu_1,\ldots,\mu_\a\to \infty$
when the ratio of two Pfaffians is replaced by a minor 
$
\pf 
\left[{
~I_{++} \ \ \; S_{++}  \atop
-S_{++}^T\ \ D_{++} }\right]
$,
the correlation function tends
to that of Laguerre symplectic ensemble
computed by Nagao and Forrester,
Eqs.(2.27), (2.25), and (4.7$\sim$9) in Ref.\cite{NF95}
(whose notations are related to ours via an unfolding
change of variables
\begin{eqnarray}
&&
I_{++}(\z,\z')=
- I_4\bigl(\frac{\zeta^2}{8N}, \frac{{\zeta'}^2}{8N}\bigr)
,\nonumber\\
&&
S_{++}(\z,\z')=
- \frac{1}{8N}S_4\bigl(\frac{\zeta^2}{8N}, \frac{{\zeta'}^2}{8N}\bigr)
,
\label{55}\\
&&
D_{++}(\z,\z')=
\frac{1}{(8N)^2}
D_4\bigl(\frac{\zeta^2}{8N}, \frac{{\zeta'}^2}{8N}\bigr)
,\nonumber
\end{eqnarray}
and $\nu=2a-1/2$).
It is easy to confirm that
the correlation functions
satisfies the decoupling sequence (\ref{decouple})
as each of the masses are sent to 
infinity.
To illustrate this decoupling, we exhibit in Fig.2 a plot of
the spectral density $\rho_s(\z;\m,\m)$ ($\n=0, p=1, \a=1$)
that interpolates between known results for
the chiral and quenched limits.

After the Authors announced the above formula in \cite{NN}, 
Akemann and Kanzieper \cite{AK} presented 
another form of the asymptotic 
correlations in a special case of quadruply 
degenerate masses. In Appendix C, we shall reproduce 
their result as a confluent limit of our formula.  
\acknowledgments
This work was supported in part (SMN) by
JSPS Research Fellowships for Young Scientists, and
by Grant-in-Aid No.\ 411044 
from the Ministry of Education, Science, and Culture, Japan.

\appendix
\setcounter{equation}{0}
\section{quaternion determinant}
A quaternion is defined as a linear combination of 
four basic units $\{1, e_1, e_2, e_3 \}$:
\begin{equation} 
q=q_0+{\bf q} \cdot {\bf e}=q_0+q_1e_1+q_2e_2+q_3e_3. 
\end{equation}
Here the coefficients $q_0, q_1, q_2$ and $q_3$ are real 
or complex numbers. The first part $q_0$ is called the scalar 
part of $q$. The quaternion basic units satisfy 
the multiplication laws 
\bea 
&& 1 \cdot 1=1,\ 1 \cdot e_j=e_j \cdot 1=e_j,\ j=1,2,3, 
\nonumber\\
&&e_1^2=e_2^2=e_3^2=e_1e_2e_3=-1. 
\eea
The multiplication is associative and in general not commutative. 
The dual ${\hat q}$ of a quaternion $q$ is defined as 
\begin{equation} 
{\hat q} = q_0-{\bf q} \cdot {\bf e}. 
\end{equation} 
For a selfdual $N \times N$ matrix $Q$
with quaternion elements $q_{jk}$ has a dual 
matrix ${\hat Q}=[{\hat q}_{kj}]$. 
The quaternion units can be represented 
as $2 \times 2$ matrices
\bea 
&&1 \rightarrow 
\left[ \begin{array}{cc} 1 & 0 \\ 0 & 1 \end{array} 
\right], \;\; 
e_1 \rightarrow \left[ \begin{array}{cc} 0 & -1 \\ 1 & 0 \end{array} 
\right],
\nonumber\\
&&
e_2 \rightarrow \left[ \begin{array}{cc} 0 & -i \\ -i & 0 \end{array} \right], 
\;\; 
e_3 \rightarrow \left[ \begin{array}{cc} i & 0 \\ 0 & -i \end{array} \right]. 
\eea
We define a quaternion determinant {\rm Tdet} of a 
selfdual $Q$ (i.e., $Q={\hat Q}$) as
\begin{equation} {\rm Tdet}\, Q = \sum_P (-1)^{N-l} \prod_1^l (q_{ab} 
q_{bc} \cdots q_{da})_0, 
\end{equation}
where $P$ denotes any permutation of the indices 
$(1,2,\ldots,N)$ consisting of $l$ exclusive cycles of the form 
$(a \rightarrow b \rightarrow c \rightarrow \cdots \rightarrow d \rightarrow a)$ 
and $(-1)^{N-l}$ is the parity of $P$. The subscript $0$ means that 
the scalar part of the product is taken over each cycle. Note that a 
quaternion determinant of a selfdual quaternion matrix is always a scalar. 
The quaternion determinant can as well be represented
by the $2 N \times 2 N$ representation $C(Q)$ \cite{DysonQ}:
\begin{equation}
{\rm Tdet}Q = {\rm Pf}[{\bf J} C(Q)], \ \ \  
{\bf J} = \openone_N \otimes \left[ \begin{array}{cc} 
0 & 1 \\ -1 & 0 \end{array} \right].
\label{Dysoneq}
\end{equation}

\setcounter{equation}{0}
\section{quaternion kernel for the orthogonal ensemble}
Recently Akemann and Kanzieper \cite{AK} derived 
the asymptotic correlation function in a special case 
$p=\alpha = 1$. Though their result is clearly in 
agreement with that special case of ours,  
there is a difference in the appearance because 
they adopted alternative asymptotic formulas. 
Their formulas are based on an asymptotic relation 
for $z, z' >0$ and $\nu \neq 1$ (see below): 
\bea
S(z,z') &\sim& 2 N \left[
2\int_0^1 dt\,t\, 
J_{\nu-1}(t\z) J_{\nu-1}(t\z') 
\right.
\label{B1}\\
&&
\left.
-  \frac{J_{\nu}(\z')}{\z'} 
\left( \int_0^{\z} ds\, J_{\nu-2}(s) - 1 \right)
\right], 
\nonumber
\eea
where we have adopted the microscopic variables (\ref{zz}).
This asymptotic relation 
was derived by Forrester, Nagao, and Honner \cite{FNH} in a 
study of parametric 
random matrix ensembles. 
Note that the first integral in the above
is equal to the
Bessel kernel,
\beq
\frac{\z J_{\n}(\z) J_{\n-1}(\z')- J_{\n-1}(\z)\z' 
J_{\n}(\z')}{\z^2-{\z'}^2}.
\eeq
From the derivation the equivalence of (\ref{Szz}) and (\ref{B1}) 
was well established. However, it is worth directly proving it here 
for an unambiguous identification. 

Using the Bessel function identities, we can readily see that 
\widetext
\Lrule
\renewcommand{\theequation}{B\arabic{equation}}
\begin{eqnarray}
&&{t} \int_0^{\z } ds\, s\, 
J_{\nu-1}({t} s)  
 =  \z\,  J_{\nu}({t}\z ) + 
(\nu-1) \int_0^{\z } ds\, J_{\nu}({t} s), 
\nonumber \\   
&&{t} \int_0^{\z}   ds\,
J_{\nu}({t} s) =  
- J_{\nu-1}({t}\z ) +  J_{\nu-1}(0) + (\nu-1) 
\int_0^{\z }  \frac{ds}{s} 
J_{\nu-1}({t} s). 
\label{B2}
\end{eqnarray}
Substitution of Eq.(\ref{B2}) into (\ref{Szz}) yields
\begin{eqnarray}
S(z,z') & \sim & 
2N\left[
\frac{\z}{\z'} \int_0^1 dt\,t\, J_{\nu}({t}\z ) 
J_{\nu}({t}\z')  
 +  
\int_0^1 dt\,t \left( J_{\nu-1}({t}\z ) - J_{\nu-1}(0) \right)
J_{\nu-1}({t}\z') 
+ \frac{J_{\nu}(\z' )}{\z' }  \right.
\nonumber \\ 
&  &+
\left.
\frac{\nu-1}{\z' } \int_0^1 d t\,t 
\int_0^{\z } d s \,J_{\nu}({t} s) J_{\nu}({t}\z'   ) 
-  (\nu-1) \int_0^1 d t\,t 
\int_0^{\z } \frac{d s }{s} J_{\nu-1}({t} s) 
J_{\nu-1}({t}\z' ) \right]. 
\label{B4}
\end{eqnarray}
By partial integrations, we find
\begin{equation}
    \frac{\z}{\z'} \int_0^1 d t\,t\, J_{\nu}({t}\z ) J_{\nu}({t}\z'   ) 
 = \int_0^1 d t\,t\, J_{\nu-1}({t}\z ) J_{\nu-1}({t}\z'   ) 
-  J_{\nu-1}(\z ) \frac{J_{\nu}(\z' )}{\z' } 
\label{B5}
\end{equation}
and
\beq
 \frac{1}{\z' } \int_0^1 d t\,t \int_0^{\z } d s \,
J_{\nu}({t} s) J_{\nu}({t}\z'   )
=  \int_0^1 d t\,t \int_0^{\z }
\frac{ d s }{s} J_{\nu-1}({t} s) J_{\nu-1}({t}\z'   ) 
  - \int_0^{\zeta} \frac{{d} s}{s} 
J_{\nu - 1}(s) \frac{J_{\nu}(\zeta')}{\zeta'}.
\label{B6}
\eeq
We substitute (\ref{B5}) and (\ref{B6}) into (\ref{B4}) and obtain
\begin{eqnarray}
&&S(z,z')  \sim\\  
&&2N\left[
 2  \int_0^1 d t\,t\, J_{\nu-1}({t}\z ) J_{\nu-1}({t}\z')  
 -  J_{\nu-1}(0) \int_0^1 d t\,t\, J_{\nu-1}({t}\z') 
 -   \left( J_{\nu-1}(\z ) + (\nu-1) \int_0^{\z } \frac{d s}{s} 
{J_{\nu-1}(s)} - 1 \right) \frac{J_\nu (\z')}{\z' }
\right] . \nonumber
\end{eqnarray}
Again by a partial integration, we find 
\begin{equation}
J_{\nu-1}(\z ) + (\nu-1) \int_0^{\z }
\frac{ d s }{s} J_{\nu-1}(s)
= J_{\nu-1}(0) + \int_0^{\z } ds\, J_{\nu-2}(s)  
\end{equation}
and thus arrive at the desired result (\ref{B1}) provided that $\nu$ is not 
equal to $1$. Since $J_{\nu-1}(0) = \delta_{\nu,1}$, the case $\nu = 1$ 
is exceptional. Ä

Similarly, for $z<0, z'>0$ and $\nu \neq 1$, an alternative expression 
\begin{equation}
S(z',z) \sim  2N\left[  
2\int_0^1  dt\,t \,J_{\nu-1}(t\z) 
I_{\nu-1}(t\m)
-   \frac{I_{\nu}(\m)}{\m} \left( \int_0^{\z}ds\,J_{\nu-2}(s) 
  - 1 \right) \right]
\end{equation}
\Rrule
\narrowtext
\noindent
in terms of the microscopic variables (\ref{mz})
is available.
\renewcommand{\thesection}{\Alph{section}}
\renewcommand{\theequation}{C\arabic{equation}}
\setcounter{equation}{0}
\section{quaternion kernel for the symplectic ensemble} 
In previous \cite{NN} and this works, the Authors evaluated Dirac eigenvalue 
correlation
functions 
for the symplectic ensemble with doubly degenerate 
masses. If masses are quadruply degenerate, the evaluation of the 
correlation functions is easier 
because the conventional integration method for the massless 
Laguerre ensemble works without any modification. 
In that case, the multiple integral we need to evaluate is  
\begin{eqnarray}
& & \Xi_p(x_1,\ldots,x_p;\{m\}) \nonumber \\ 
& = & \frac{1}{(N-p)!} 
\int_0^{\infty} \!\! \cdots \int_0^{\infty}
\prod_{j=p+1}^N dz_j
\prod_{j=1}^N x_j^{2 \nu + 1} {\rm e}^{- 4 x_j} \\
&&\times \prod_{j=1}^N  
\prod_{i=1}^{\alpha} (x_j +m_i^2)^4 \prod_{j > k}^N
\mid x_j - x_k \mid^4 ,\nonumber 
\end{eqnarray} 
where we set $N_f=4\a$ and
\[
\{m\}=(\overbrace{m_1,\ldots,m_1}^{4},\ldots,\overbrace{m_\a,\ldots,m_\a}^{4}).
\]
The conventional `massless' theory \cite{MM,NF95,NF98,NF99} tells us that the 
correlation functions are written in terms of 
quaternion determinants:
\bea
&&\sigma(x_1,\ldots,x_p;\{m\}) 
=  \frac{\Xi_p(z_1,\ldots,z_{\a+p})}{\Xi_0(z_1,\ldots,z_{\alpha})} \nonumber \\ 
&&=   \frac{{\rm Tdet}[f(z_j,z_k)]_{j,k=1,\ldots,\alpha+p}}{{\rm 
Tdet}[f(z_j,z_k)]_{j,k=1,\ldots,\alpha}},
\end{eqnarray}
where we have adopted the notation (\ref{zj}).

The quaternion function $f(z,z')$ is represented as
\begin{equation}
f(z,z')= \left[ \begin{array}{cc} S(z,z') & I(z,z') \\ 
D(z,z') & S(z',z) \end{array} \right].
\end{equation}
We evaluate the asymptotic limit of the 
quaternion function $f(z,z')$ in each of the three cases 
\[
(a)\ z, z'>0,\ \ \  (b)\ z<0, z'>0,\ \ \ (c)\ z, z'<0, 
\]
as in Sect.2. in terms of 
the
microscopic variables $\zeta, \zeta', \m, \m'$ 
defined by Eqs.(\ref{zz}), (\ref{mz}), (\ref{mm}), respectively.
\par
\bigskip
\noindent
$(a)$ $z, z'>0$
\par
\bigskip
\noindent
In the case $z, z'>0$, Nagao and Forrester  \cite{NF95} 
derived the asymptotic limit (\ref{55}), that is
\bea
\frac{1}{8N}S(z,z') &\sim& -S_{++}(\zeta, \zeta'),\nonumber\\
\frac{1}{(8N)^2}D(z,z') &\sim& D_{++}(\zeta, \zeta'),
\label{C5}\\
I(z,z') &\sim& -I_{++}(\zeta, \zeta'),\nonumber
\eea
where $S_{++}$,  $D_{++}$,  and $I_{++}$ are
defined in Eq.(\ref{SDI}).
\par
\bigskip
\noindent
$(b)$ $z<0, z'>0$
\par
\bigskip
\noindent
Using the asymptotic formula for the Bessel function (\ref{LJI}), 
we can similarly treat negative argument cases to obtain
\widetext
\Lrule
\renewcommand{\theequation}{C\arabic{equation}}
\bea
\frac{1}{8N}S(z,z') &\sim& -S_{-+}(\mu, \zeta),\nonumber\\
\frac{1}{8N}
S(z',z) &\sim&  \z \int_0^1 dt\, {t}^2 \int_0^1 du  \bigl(
{J_{2 \nu}(2 {t u}\z ) } I_{2 \nu+1}(2 t \m) - 
J_{2 \nu}(2 {t}\z) u\, I_{2 \nu + 1}(2 {t u}\m) \bigr) , 
\nonumber\\ 
\frac{1}{(8N)^2}D(z,z') &\sim& -
\int_0^1 dt\,t^3 \int_0^1 du\,u  \bigl( 
I_{2 \nu + 1}(2 {t u}\m ) J_{2 \nu + 1}(2 {t}\z   ) - 
I_{2 \nu + 1}(2 {t}\m  ) J_{2 \nu + 1}(2 {t u}\z)  \bigr) ,
\\
I(z,z') &\sim& -I_{-+}(\m, \zeta),\nonumber
\eea
where $S_{-+}$ and $I_{-+}$ are
defined in Eq.(\ref{SDI}).
\par
\bigskip
\noindent
$(c)$ $z, z'<0$
\bea
\frac{1}{8N}S(z,z') &\sim& 
\m \int_0^1 dt\, {t}^2 \int_0^1 du \bigl(
{I_{2 \nu}(2 {t u}\m ) } I_{2 \nu+1}(2 {t}\m') - 
I_{2 \nu}(2 {t} \m) u\,I_{2 \nu + 1}(2 {t u}\m' ) \bigr), 
\nonumber\\
\frac{1}{(8N)^2}D(z,z') &\sim& 
- \int_0^1 dt\,t^3 \int_0^1 du\,u \bigl(
I_{2 \nu + 1}(2 {t u}\m ) I_{2 \nu + 1}(2 {t}\m' ) - 
I_{2 \nu + 1}(2 {t}\m  ) I_{2 \nu + 1}(2 {t u}\m' )  
\bigr) , \\
I(z,z') &\sim& -I_{--}(\m, \m'),\nonumber
\eea
\Rrule
\narrowtext
\renewcommand{\theequation}{C\arabic{equation}} 
\noindent
where $I_{++}$ is
defined in Eq.(\ref{SDI}).

We can use 
Dyson's equality (\ref{Dysoneq}) to see that the above quaternion determinant 
expression is identical to the limit of quadruple mass degeneracy of the 
general formula (\ref{Pf4even}) employing Pfaffians.
In this case, 
yet another equivalent asymptotic 
formula was recently presented by Akemann and Kanzieper \cite{AK}. 
Now we shall directly demonstrate the equivalence. 
We should firstly note that, under the change 
of the quaternion elements
\begin{eqnarray}
S(z,z') & \rightarrow & {\tilde S}(z,z') \equiv S(z,z') -
\frac{1}{W(z')} \frac{dW(z')}{dz'} 
I(z,z'), \nonumber \\ 
I(z,z') & \rightarrow & {\tilde I}(z,z') \equiv 
- \int_z^{z'} {\tilde S}(z,z'') dz'' , \\   
D(z,z') & \rightarrow & {\tilde D}(z,z')
\equiv \frac{\partial}{\partial z} 
{\tilde S}(z,z'), \nonumber
\end{eqnarray}
where
\begin{equation}
W(z) = |z|^{\nu + {1}/{2}} {\rm e}^{- 2 z}.
\end{equation}
the quaternion determinant is unchanged. This transformation was 
introduced in Ref. \cite{NW}. 

For $z, z'>0$, we find an identity
\widetext
\Lrule
\renewcommand{\theequation}{C\arabic{equation}} 
\beq
{\z\z'} \int_0^1 dt\,t \int_0^1 {du}
J_{2 \nu}(2 {t u}\z   ) J_{2 \nu}(2 {t}\z'   )  
=  {\z'}^{2} \int_0^1 dt\,t \int_0^1  \frac{du}{u} 
J_{2 \nu + 1}(2 {t u}\z   ) J_{2 \nu + 1}(2 {t}\z'   ) + 
\frac{\z' }{2} J_{2 \nu}(2 \z' ) \int_0^1  \frac{du}{u} 
J_{2 \nu + 1}(2  u \z),
\label{C17}
\eeq
by a partial integration.
The asymptotic formulas
(\ref{C5}), together with 
the identity (\ref{C17}), yield
\begin{eqnarray}
{\tilde S}(z,z')
&\sim&
2N \left[
2\int_0^1 dt\,t\,  J_{2 \nu + 1}(2 {t}\z'   ) \left(
2 {t}\z  \int_0^1 {du}{} J_{2 \nu}(2 {t u}\z   ) 
- (2 \nu + 1) \int_0^1  \frac{du}{u} J_{2 \nu + 1}(2 {t u}\z   ) 
\right) \right.
\nonumber\\ 
&&
+2\frac{\z}{\z'} \int_0^1 dt\,t \,J_{2 \nu}(2 {t}\z ) 
\left( - 2 {t}\z'    \int_0^1 du\,u\, J_{2 \nu + 1}(2 {t u}\z' ) + 
(2 \nu + 1) \int_0^1 {du}\,J_{2 \nu}(2 {t u}\z' ) 
\right) 
\nonumber\\
&&
\left.
- (2 \nu + 1)
\frac{J_{2 \nu}(2 \z' ) }{\z' }  
\int_0^1 \frac{du}{u} J_{2 \nu + 1}(2 u\z) \right].
\label{C18}
\end{eqnarray}  
Partial integrations give rise to the Bessel function equalities 
\begin{eqnarray}
&&(2 \nu + 1) \int_0^1  \frac{du}{u} J_{2 \nu + 1}(2 {t u}\z   )  = 
-  J_{2 \nu + 1}(2 {t}\z ) + 2 {t}\z  \int_0^1  
{du}\, J_{2 \nu}(2 {t u}\z   ),  \nonumber
\\  
&&(2 \nu + 1) \int_0^1  {du}\, J_{2 \nu}(2 {t u}\z' ) 
=  J_{2 \nu}(2 {t}\z') + 2 {t}\z'    \int_0^1 du\,u \,
J_{2 \nu + 1}(2 {t u}\z' ).
\label{C19}
\end{eqnarray} 
Substituting (\ref{C19}) into (\ref{C18}) and using the formula 
(\ref{B5}), we obtain
\beq
{\tilde S}(z,z') 
 \sim  2N\left[
4  \int_0^1 dt\,t\,
J_{2 \nu + 1}(2 {t}\z ) J_{2 \nu + 1}(2 {t}\z'   ) 
+  \frac{J_{2 \nu}(2 \z' )}{\z' } J_{2 \nu + 1}(2 \z ) 
- (2 \nu + 1) \frac{J_{2 \nu}(2 \z' )}{\z' } \int_0^1  \frac{du}{u} 
J_{2 \nu + 1}(2 u\z) \right] .
\eeq
By a partial integration, we can rewrite it as
\begin{equation}
{\tilde S}(z,z') \sim 2N\left[
4  \int_0^1 dt\,t\, J_{2 \nu + 1}(2 t\z ) 
J_{2 \nu + 1}(2 t\z'   ) -  
\frac{J_{2 \nu}(2 \z' )}{\z' } 
 \int_0^{2 \z } ds\, J_{2 \nu + 2}(s)\right] .
\end{equation} 
\narrowtext
\noindent
This is the asymptotic formula derived by Forrester, Nagao, and Honner 
\cite{FNH} in a study of parametric random matrices and 
then applied by Akemann and Kanzieper \cite{AK} to the massive 
Dirac operator problem. Thus we established the equivalence 
in the case of quadruply degenerate masses.  
We can straightforwardly extend it to the formulas 
with $z$ and/or $z'$ negative.

\newpage
\widetext
\begin{figure}
\vspace{20mm}
\begin{center}
\leavevmode
\epsfxsize=300pt
\epsfbox{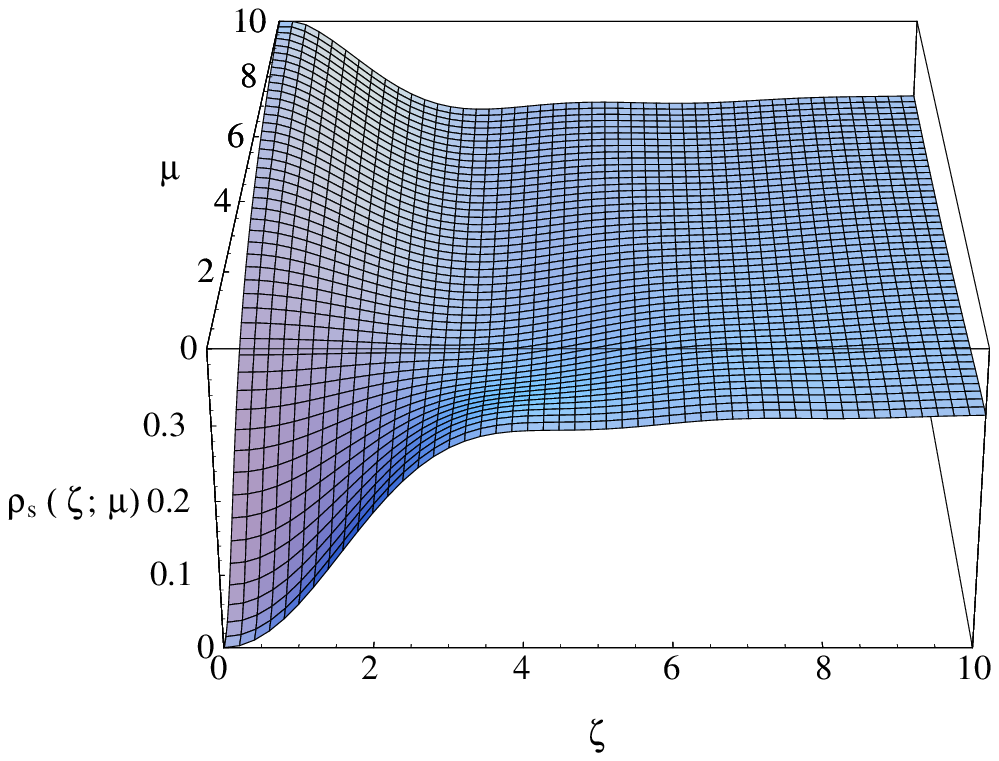}
\end{center}
\caption{The scaled spectral density $\rho_s(\z;\m)$ for the chiral 
orthogonal ensemble with one ($\alpha=1$) flavor.}
\vspace{15mm}
\begin{center}
\leavevmode
\epsfxsize=300pt
\epsfbox{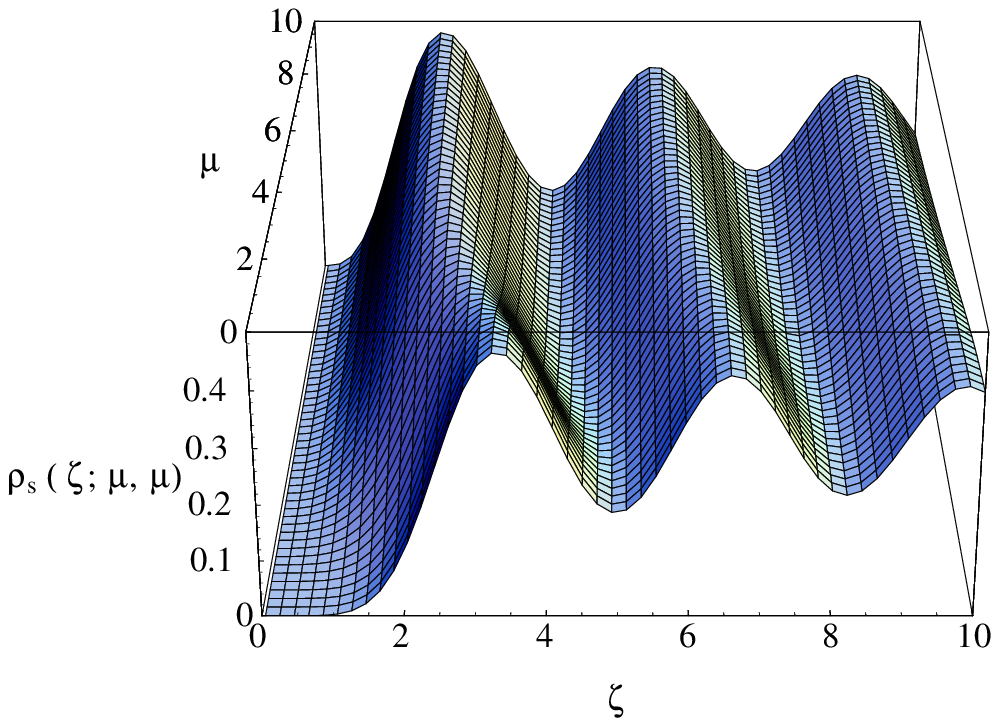}
\end{center}
\caption{The scaled spectral density $\rho_s(\z;\m,\m)$ for the chiral 
symplectic ensemble with two degenerate ($\alpha=1$) flavors.}
\end{figure}
\end{document}